\documentclass[preprint]{aastex}
\usepackage{epsf}

\newcommand{\dv}{r^{1/4}}

\newcommand{\om}{\Omega_M}
\newcommand{\ol}{\Omega_\Lambda}

\begin{document}

\slugcomment{Accepted for publication in {\it The Astronomical Journal}}

\title{A Deep Multicolor Survey. VI. Near-Infrared Observations, Selection 
Effects, and Number Counts\altaffilmark{1}}

\author{Paul Martini\altaffilmark{2}}

\affil{Department of Astronomy, 140 W. 18th Ave., Ohio State University, \\
Columbus, OH 43210}

\altaffiltext{1}{Based on observations obtained at 
 MDM Observatory, operated by Columbia University, Dartmouth College, the 
 University of Michigan, and the Ohio State University}

\altaffiltext{2}{Current Address: Carnegie Observatories, 813 Santa Barbara St., Pasadena, CA 91101, martini@ociw.edu}

\begin{abstract}

I present near-infrared $J$ ($1.25\mu$m), $H$ ($1.65\mu$m), and $K$
($2.2\mu$m) imaging observations of 185 square arcminutes in 21 high galactic
latitude fields. These observations reach limiting magnitudes of $J \sim 21$
mag, $H \sim 20$ mag and $K \sim 18.5$ mag. The detection efficiency,
photometric accuracy and selection biases as a function of integrated object
brightness, size, and profile shape are quantified in detail.
I evaluate several popular methods for measuring the integrated light of
faint galaxies and show that only aperture magnitudes provide an
unbiased measure of the integrated light that is independent of apparent
magnitude. These $J$, $H$, and $K$ counts and near-infrared colors are in best
agreement with passive galaxy formation models with at most a small
amount of merging (for $\om = 0.3$, $\ol = 0.7$).

\end{abstract}


\section{Introduction} \label{sec:intro}

The surface density of galaxies as a function of integrated brightness, also
known as the number--magnitude relation, is one of the classic
tests of observational cosmology. This test has a long, distinguished history
as a tool for studying the nature of the universe and the evolution of
galaxies \citep[e.g.][]{hubble34, sandage61, tinsley77}.
Surveys to progressively deeper apparent magnitude limits have measured the
number--magnitude relation at a range of wavelengths, and this work has been
reviewed by \citet{koo92} and \citet{ellis97}.
The observed number--magnitude relations are compared
to the predictions of galaxy evolution models, such as those by
\citet{yoshii88}, \citet{guiderdoni90}, and \citet{gardner98} in order to
determine cosmological parameters and study galaxy formation and evolution.

Near-infrared (1 -- 2.5$\mu$m, hereafter NIR) galaxy counts are a valuable
addition to studies of galaxy formation and evolution with the
number--magnitude
relation. The spectral energy distributions (SEDs) of most galaxies are
relatively constant as a function of wavelength in the NIR because
this wavelength regime is dominated by light from old, evolved stars.
The shape and uniformity of galaxy SEDs in the NIR results in
relatively small $k$-corrections for galaxies up to $z \sim 1$ and
therefore the NIR luminosity is representative of the total stellar
luminosity. A recent burst of star formation, in contrast, can significantly
increase the brightness of a galaxy at visible and $UV$ wavelengths. This may
cause a galaxy with a large amount of current
star formation to appear as bright as a more quiescent galaxy with a much
larger stellar mass. The presence of a great deal of dust will also attenuate
the brightness of galaxies at these wavelengths and further complicate the
relation between $UV$ or visible luminosity and total stellar luminosity.  NIR
number counts to a limit of $\sim 20$ mag, which are dominated by galaxies at
$z < 1$, provide a less uncertain means of evaluating models of
galaxy formation. The old, evolved stars that dominate the NIR
light are also good tracers of the total stellar mass and therefore
NIR measurements are an excellent method for direct study of the stellar
mass evolution in galaxies in the context of hierarchical galaxy formation
\citep{brinchmann00}.

NIR number counts to date have mostly been obtained at $K$
\citep{gardner93,cowie94,djorgovski95,mcleod95,gardner96,moustakas97} as
this is the longest-wavelength atmospheric transmission window with sufficient
sensitivity to detect high-redshift galaxies in a reasonable time. Faint
counts have also been obtained
at $H$ \citep{teplitz98,yan98} with NICMOS on the
{\it Hubble Space Telescope} (HST), though given its small field of view, there
are few bright galaxies in this sample. Some observations have also been
obtained at both $J$ and  $K$ \citep{bershady98,teplitz99,vaisanen00} and 
these studies
have used the mean NIR colors as a function of apparent magnitude
as an additional constraint for galaxy evolution models.

I have obtained $J$, $H$, and $K$ observations of 185 square arcminutes of
high galactic latitude fields which extend to $J \sim 21$, $H \sim 20$,
and $K \sim 18.5$. The median redshift to $K = 18.5$ is $z \sim 0.5$ in the
$K$-selected Hawaii Redshift Survey \citep{cowie96}. These observations
are made up of 21
subfields of the Deep Multicolor Survey (DMS) fields observed by
\citet{hall96a}. The DMS covers a total area of 0.83 square degrees in
six filters: $UBVRI_{75}I_{86}$. The DMS is comprised of 6 different fields
and to date this rich dataset has been used to study the quasar luminosity
function \citep{hall96b,kennefick97}, the evolution of blue galaxies
\citep{liu98}, and the M dwarf distribution, mass, and luminosity function
\citep{martini98}. The selection criteria for the DMS fields were
relatively low galactic foreground extinction and avoidance of very bright
stars. Five of the fields are equatorial to facilitate follow-up
spectroscopy from both hemispheres.

In \S \ref{sec:obs} I describe the NIR observations, while in \S \ref{sec:proc} 
I discuss the image processing, photometric
solution, and catalogs. I discuss the detection efficiencies in
\S \ref{sec:det}, galaxy photometry in \S \ref{sec:phot}, and
star/galaxy separation in \S \ref{sec:stars}. The galaxy number counts
and NIR colors of these galaxies are presented in
\S\S \ref{sec:num} and \ref{sec:colors}.
In a separate paper \citep{martini00} I combine these NIR observations
with the $UBVRI_{75}I_{86}$ CCD data to compute photometric
redshifts for this NIR sample to study the visible-NIR
colors of these galaxies, particularly the Extremely Red Objects. 

\section{Observations} \label{sec:obs}

A total of 21 randomly-selected subfields of the DMS were observed through
the NIR $J$, $H$, and $K$ filters. These images were
obtained with TIFKAM\footnote{TIFKAM: The Instrument Formerly Known As MOSAIC.
This instrument is also known as ONIS at Kitt Peak.}
\citep{pogge98}, a NIR imager/spectrograph with a 512x1024 InSb
detector on the 2.4m Hiltner telescope of the MDM Observatory. The f/7 camera
was used for all of the observations; this camera gives a plate scale of
$0.3''$ pix$^{-1}$ on the Hiltner telescope.
Table~\ref{tbl:log} is a log of the observations, listing the date each field
was observed in each filter and the seeing measured in that field/filter
combination.
Each of the 21 fields was observed for a total of 60 minutes on-source
integration time per filter. The observations were obtained in a six-point
dither pattern with a 10$''$ -- 15$''$ offset between positions. Individual
exposures in the dither pattern were one to two minutes and comprised a
combination of integration time and coadds such that the sky level in the
frames was less than 10000 counts. Tests of the detector linearity showed that
the array is less than 1\% nonlinear at this count level.
Dark frames were taken nearly every night with the same combination of
exposure time and number of coadds as the science frames.

Standard stars from \citet{persson98} were observed each night at a range of
airmass. A series of red stars was also observed to determine the
transformation of the instrumental magnitude system to the CTIO/CIT photometric
system \citep{elias82}. Red stars were observed at least one night per
observing run as there were several changes to the filters
over the course of the observations for this program. These changes are
discussed below in \S \ref{sec:stand}. The standard
stars were observed in a five or six point dither pattern with 15$''$ offsets,
though with only 15 -- 30 seconds on-source integration per dither position.

\section{Image Processing} \label{sec:proc}

\subsection{Survey Fields}

All of the data were reduced with IRAF\footnote{IRAF is distributed by the
National Optical Astronomy Observatories, which are operated by the
Association of Universities for Research in Astronomy, Inc., under cooperative
agreement with the National Science Foundation.}
and many of the data processing steps took advantage of the
PHIIRS\footnote{Available at http://iraf.noao.edu/iraf/web/contrib.html}
package, a collection of IRAF routines compiled by Pat Hall. The first step
in the data processing was subtraction of a dark frame from each image. The
individual images were then flat-fielded and sky-subtracted using
``running'' flat and sky frames. The running flats were created by averaging
the six neighboring images using a percentile clipping algorithm to remove
bright objects. This image was then normalized to unity to create a skyflat
for each image. The running sky frames were similarly created by averaging
the six neighboring frames. At this stage the images were inspected and
images with anomalously high noise or serious tracking/guiding errors
(generally two to three per field) were removed from further processing.
The good images
were then shifted and coadded together using an offset table of integer shifts
to create a final mosaic image. This image was used to identify objects
and, together with the offset table, was used to make an object mask for each
of the individual frames. The flat field and sky subtraction steps were
then repeated using these individual object masks.

TIFKAM has a noise pattern in the lower half of the array due to vibration
of the mechanical cryocooler. This pattern is a few counts in amplitude
and runs along rows in the form of a sine wave with a period six times the
width of the array. Because the period is much longer than the array width
and the amplitude is less than the sky noise, simply subtracting the average of
each row, after masking out objects, removes most of this feature.
After the cryocooler pattern was subtracted, the
individual images were shifted and added into the final mosaic. This region
was then trimmed to only include the area of the sky present in more than
83\% of the individual frames. As the offsets were generally small and
consistent, this step produced a final mosaic of relatively constant noise
with minimal loss of field.
Several of the fields in Table~\ref{tbl:log} were initially observed under
nonphotometric conditions. Calibration images of these fields were
obtained on photometric nights and these images were reduced in the same
manner as above. Three to six bright, stellar objects were
used to calibrate the fields observed under nonphotometric conditions.

\subsection{Standards} \label{sec:stand}

The standard stars were processed in an identical manner to that outlined
above, though with two exceptions. First, the cryocooler pattern was not
subtracted from these frames. Tests of the images with artificial stars
showed that this pattern did not affect the photometry of bright
stars. Secondly, the five to six observations
of each standard were photometered individually, rather than one measurement of
a shifted and added frame. The standard stars were measured with PHOT using a
$15''$ radius aperture and a mean sky measured in an annulus extending from
$25''$ to $35''$ radius.

To convert the instrumental magnitudes to the CTIO/CIT system, the
following equations were solved:
\begin{equation}
\begin{array}{c}
J = j + j_0 + j_1\;(X - 1.3) + j_2\;(J-K - 1.4) \\
H = h + h_0 + h_1\;(X - 1.3) + h_2\;(J-K - 1.4) \\
K = k + k_0 + k_1\;(X - 1.3) + k_2\;(J-K - 1.4)
\end{array}
\end{equation} \label{eqn:phot}
where $j,h,k$ are the instrumental magnitudes, $j_0,h_0,k_0$ are the
photometric zeropoints, $j_1,h_1,k_1$ are the airmass coefficients, and
$j_2,h_2,k_2$ are the color terms. $X = 1.3$ is the
mean airmass of the fields observed in this survey and $J-K = 1.4$
is approximately the mean $J-K$ color of galaxies in the range $K = 16 - 19$
reported by \citet{saracco99}. While the photometric zeropoints and
airmass coefficients may vary from night to night, the color terms should
not vary so long as the filters remain unchanged. To take advantage of
the observations of red stars over several nights, I therefore performed a
multidimensional fit \citep[e.g.][]{gould95} to all of the
nights of a given observing run simultaneously. For each filter, this allowed
the zeropoint and airmass coefficient to vary but not the color term.  Over a
given observing run, the photometric zeropoint varied by a few percent. The
airmass coefficient was more variable, with large changes generally temporally
correlated with changes in the weather
\citep[as discussed by][]{frogel98}.
If standards at large airmass were not observed on a particular night,
the average coefficient of the temporally adjacent nights on that observing
run was used.  Most of the fields were observed at $X < 1.5$ (and all at less
than two) so uncertainties in the airmass coefficients translate to less than
one percent uncertainties in the photometry.  All of the photometric solution
coefficients are listed in Table~\ref{tbl:phot}, along with the
root-mean-square variation in these solutions.

As mentioned above, there were several changes to the filters in TIFKAM
over the course of this project. The $J$ filter used in 1997 October
had considerable structure in the wings of the point spread function (PSF)
and therefore it was replaced before the 1998 April observing run. The
new $J$ filter had a red leak such that the measured sky brightness was
$\Sigma_J \approx 13$ mag arcsec$^{-2}$, compared to measurements of
$\Sigma_J \approx 15.5 - 16$ mag arcsec$^{-2}$ on the other observing runs.
After 1998 April this $J$ filter was only used in conjunction with a piece of
PK50 glass in the prefilter wheel, which serves as a red-blocking filter.
PK50 is opaque longward of $2.7 \mu$m, but transparent at shorter wavelengths.
A piece of PK50 was also added to the same filter cell as the $H$ filter
between the 1998 April and 1998 September observing runs due to the possible
presence of a slight red leak in this
filter.  Because of the changes in the $J$ and $H$ filters, the photometric
solution has three separate color terms for the $J$ filter and two
separate color terms for the $H$ filter.  The $K$ filter was unchanged over
the course of this observing program and therefore only one color term was
sought in this equation.

\subsection{Object Catalogs}

I cataloged the positions of all objects in these fields with the
SExtractor package \citep{bertin96} and the default object detection
filter. This choice was motivated by the data reduction procedure, which
creates images with ``dead'' regions, corresponding to areas with low exposure
times and higher sky noise in the final shift-and-add step. The weight map
option in SExtractor is an efficient way to insure detections on only the
useful part of each image file. In addition, the noise varies somewhat over
our images, and specifically the northern half of the array
is on average noisier than the southern half. I compared SExtractor to FOCAS
\citep{jarvis81,valdes82} and found that SExtractor did a better job of
detecting all of the objects in the fields without introducing significant
spurious detections.

\section{Detection Efficiencies} \label{sec:det}

The limiting magnitude of a survey is typically expressed as the apparent
magnitude where the probability of detecting an object is equal to
some percentage, commonly 50\% or 90\%. To obtain an accurate census of
all objects to such limiting magnitudes, one can use an estimate of
the incompleteness vs.\ apparent magnitude to correct the observed number of
objects to the true value. This simple characterization of the detection
efficiency is sufficient for the study of objects that all have the same
size and surface brightness profile and is simplest to apply to
point sources such as stars or quasars.  I discuss the related issue of
star--galaxy separation below in \S \ref{sec:stars}.

In studies of galaxies, more compact and higher surface brightness objects
are generally easier to detect. Not taking the relative ``visibility''
\citep{disney76,disney83,phillipps90,davies90} of different types of
objects into account can lead to erroneous conclusions about the intrinsic
distribution of observed galaxy properties.  The detection efficiency of a
survey should therefore be characterized as not just a function of apparent
magnitude, but also as a function of surface brightness profile and angular
size. That is, surveys should take into account the multivariate nature of
the galaxy population, rather than collapsing variations in profile shape,
size, and surface brightness into just a dependence on integrated brightness.
\citet{bershady98}, for example, separate objects
into stars, small, and large galaxies and calculate the detection efficiency
of each class separately.

The surface brightness profile of galaxies is commonly parameterized by
either an exponential profile
\begin{equation}
\Sigma(r) = {\rm exp} \left( -1.673 \frac{r}{r_h} \right)
\end{equation} \label{eqn:expdisk}
or a de~Vaucouleurs ($\dv$) profile
\begin{equation}
\Sigma(r) = {\rm exp} \left( -7.67 \left[ \left( \frac{r}{r_h}\right)^{1/4} - 1 \right] \right),
\end{equation}\label{eqn:devauc}
where $\Sigma(r)$ is the surface brightness at radius $r$ and $r_h$ is
the scale radius for each profile that encloses half of the total light.
Most galaxies can be fit by either one of these profiles or, for disk galaxies
with a bulge component, a superposition of both. These parameterizations are
useful for examining how the detection efficiency varies as a function of
galaxy size and profile shape.

The sensitivity of each of the fields in this survey is different due to
variations in the atmospheric conditions and instrument setup.
The likelihood of detecting an object of a given brightness,
size, and profile will vary from field to field and can not be accurately
represented by a single number for the entire survey, even for stellar
objects. To characterize the detection limits of this survey, I added
artificial objects produced with the IRAF ARTDATA package to each of the
fields in each filter and measured the fraction of recovered objects with
SExtractor. In addition to stellar profiles, I added galaxies with both
exponential and $\dv$ profiles with $r_h$ equal to
0.25$''$, 0.5$''$, 0.75$''$, and 1$''$ over a range of apparent magnitude.
Studies of the HST Medium Deep Survey \citep{roche97}, the
NICMOS parallel survey \citep{teplitz98,yan98}, and HST imaging of the CFRS
and LDSS galaxies \citep{lilly98} have shown that nearly all galaxies to
the apparent magnitude limits of this survey are well fit by
either exponential or $\dv$ profiles and most will have $r_h < 1''$ and
exponential profiles. An alternative to using artificial objects to measure
the detection efficiency is to extract bright objects from the survey fields,
artificially dim them, place them at random back into the original images,
and attempt to recover them. However, as the median galaxy size declines at
fainter apparent magnitudes, this may lead to an underestimate of the
detection efficiency.

I convolved the model galaxy profiles used in this analysis with a
\citet{moffat69} profile as this profile is a much better fit to the observed
profiles of bright stars than a Gaussian profile. A Moffat profile has the
functional form
\begin{equation}
I(r) = \left[1 + \left(r/\alpha\right)^2 \right]^\beta .
\end{equation} \label{eqn:moffat}
After fitting Moffat profiles to bright stars in a number of fields and filters
that spanned the range of PSF sizes, I fixed $\beta$ to be 2.5. The parameter
$\beta$ determines the strength of the power-law tail of the radial intensity
distribution, while $\alpha$ parameterizes the width of the profile peak and
is similar to $\sigma$ in a Gaussian function.
This procedure set the convolution kernel for the model galaxy profiles in
ARTDATA for each image and takes the variation in PSF from image to image into
account in the determination of the detection efficiency. The functional
form of the Moffat profile used by the ARTDATA package is slightly different
from equation~\ref{eqn:moffat}, but MOFFAT profiles produced by ARTDATA are
in reasonable agreement with this standard form. The default dynamic
range of the convolution kernel in ARTDATA for galaxy profiles
(ARTDATA.DYNRANG) is only ten, which causes an artificial truncation in galaxy
profiles.  I changed the value of this parameter to 10000, which matches the
dynamic
range of the PSF computation. While this is more computationally
intensive, it is not prohibitive.

\begin{figure*}[t]
\centerline{
\epsfxsize=3.5truein\epsfbox[65 165 550 730]{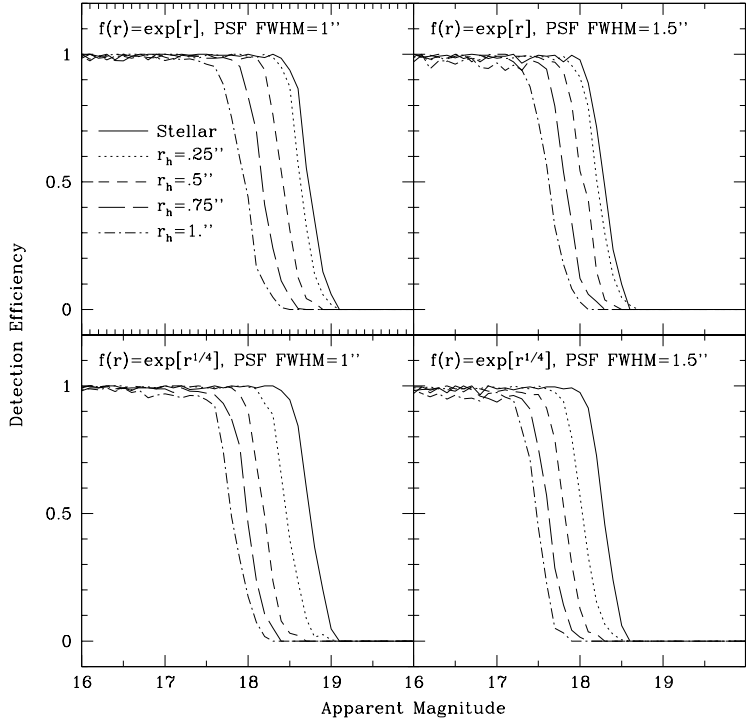}}
\caption{
\footnotesize
Variation in the detection efficiency vs.\ apparent magnitude and source
size. The upper left panel shows the detection efficiency vs.\ apparent
magnitude for a stellar profile (PSF FWHM = 1$''$ and Moffat $\beta =
2.5$) and four exponential disks with intrinsic half-light radii of
$r_h = 0.25'', 0.5'', 0.75''$, and $1''$.  The exponential disks have all been
convolved with the 1$''$ PSF. The lower left panel is the same except that
the intrinsic galaxy surface brightness profile is a de~Vaucouleur model.
The two right panels are from the same image, except with
the PSF FWHM = 1.5$''$. The sky noise and magnitude zeropoint in 
this figure is from one of the $K$ band frames, although the relative 
detection efficiencies for different sizes, profiles, and seeing are 
independent of the actual filter used. 
} \label{fig:det}
\end{figure*}

In Figure~\ref{fig:det} I show the {\it relative} detection efficiency for
these two profiles at a range of $r_h$ in a field of fixed sky brightness.
The sky noise and magnitude zeropoint in this figure is from one of the 
$K$ band frames, although the relative detection efficiencies for different 
sizes, profiles, and seeing are independent of the actual filter used. 
Throughout this paper I will use $r_h$ to parameterize the angular half-light
radius, rather than associate it with a physical scale.
The top two panels of Figure~\ref{fig:det} show an exponential disk model and
the bottom two panels show a $\dv$ profile. The left panels correspond to a
seeing FWHM of $1''$ and the right panels correspond to $1.5''$ seeing.
This figure shows that the detection efficiency for stellar objects is 0.5
-- 1 mag fainter than the detection efficiency for marginally resolved
galaxies and that the detection efficiency is similar for exponential disks
and $\dv$ profiles with the same $r_h$.
As angular size and surface brightness vary with redshift, the detection
efficiency for different classes of objects can be even more important for
selecting galaxies for a redshift survey that will be used to compute the
luminosity function \citep{dalcanton98a,petrosian98}.

\begin{figure*}[t]
\centerline{
\epsfxsize=3.5truein\epsfbox[65 165 550 730]{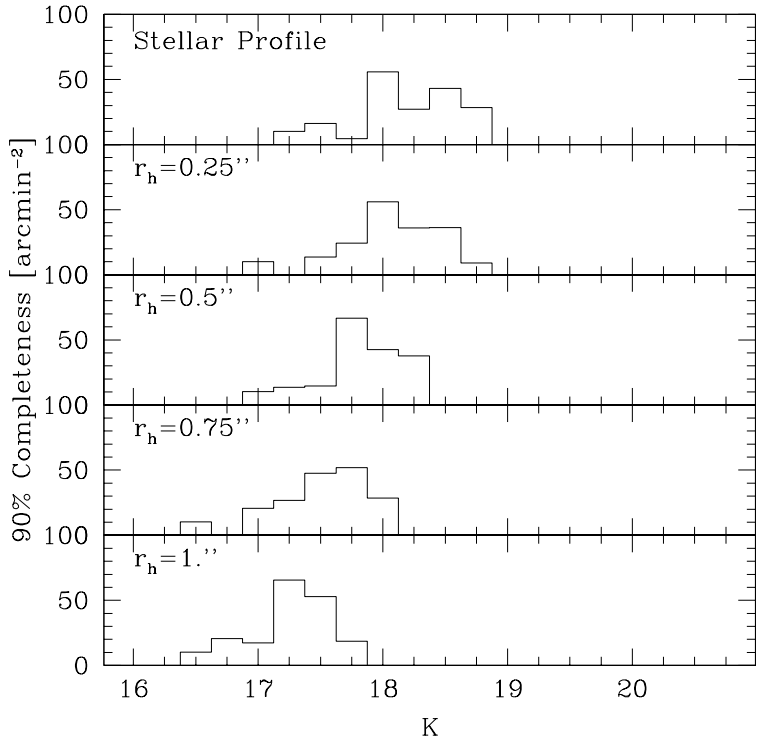}}
\caption{
\footnotesize
Histogram of survey area with a given 90\% limiting $K$ magnitude.
Each 0.25 magnitude bin shows the number of square arcminutes which are
90\% complete to that $K$ apparent magnitude for the given object size.
The top panel shows the distribution for stellar profiles,
while the remaining panels show the distribution for exponential disks
with half-light radii of $r_h = 0.25'', 0.5'', 0.75''$, and $1''$.
} \label{fig:k90}
\end{figure*}

Figure~\ref{fig:k90} illustrates the variation in detection efficiency
from field to field of this survey as a function of object size. The histogram
represents the number of square arcminutes with a given 90\% completeness limit
vs.\ $K$ magnitude for the 185 square arcminutes of this survey. The top panel
shows the distribution for stellar profiles, while the lower panels show
exponential disks with $r_h = 0.25''$, 0.5$''$, 0.75$''$, and 1$''$.
The average, area-weighted 50\% completeness limits of the survey are 
$J = 20.5$, $H = 19.5$, and $K = 18$ mag for exponential disks with 
$r_h = 0.75''$. For stellar objects these limits are $J = 21$, $H = 20$, 
and $K = 18.5$ mag. 

\section{Photometry} \label{sec:phot}

The unknown intrinsic morphology of galaxies makes galaxy photometry
considerably more challenging than measuring the apparent magnitude of a
star.  In this paper, the goal is to measure the
total, integrated brightness of galaxies to compare to models
of galaxy surface density as a function of integrated brightness. Most of the
galaxies in this sample have small angular sizes, where small in this context
corresponds to $r_h$ less than or equal to the seeing FWHM.
Due to the finite signal-to-noise ratio and the lack of numerous bright stars
in each field to exactly determine the PSF, the true morphology of the
galaxies can not be accurately deconvolved. Photometric techniques for
marginally-resolved galaxies include aperture magnitudes, isophotal
magnitudes, metric magnitudes, and various slight modifications of these
techniques.

Aperture magnitudes are the simplest to define as they just involve computing
the total flux within some given size aperture. The problem with aperture
magnitudes, however, is that they will always exclude some fraction of a
galaxy's light unless the aperture is made very large. While using a
very large aperture is a way of avoiding systematic errors in galaxy
photometry, it is impractical to apply to observations within several
magnitudes of the survey limit due to nonnegligible sky noise. A pragmatic
approach is therefore to measure the brightness of the galaxy within some
small aperture and then to correct this aperture magnitude to compensate for
the lost light, usually based on some measure of the galaxy size
\citep{glazebrook94,soifer94,cowie94,bershady98,saracco99}. For any size
aperture, the fraction of lost light will depend on both the scale size and
surface brightness profile of the galaxy. These quantities are difficult
to measure with great accuracy for galaxies near their detection limit.

An isophotal magnitude is measured by summing all of a galaxy's flux out to
some limiting isophote, usually the limiting isophotal magnitude of the survey.
A variation of the isophotal magnitude called the ``TOTAL'' magnitude is more
commonly used \citep{mcleod95,hall98,minezaki98b} and is one of the quantities
calculated by FOCAS.
The FOCAS TOTAL magnitude is defined as the sum of the flux within twice the
isophotal area, rather than just the isophotal area. The isophotal
magnitude is well-known to be biased towards underestimating the flux of all
but very bright and high surface-brightness galaxies. Using the FOCAS TOTAL
magnitude decreases the tendency of isophotal magnitudes to
underestimate galaxy size and integrated brightness by
effectively increasing the aperture size, but it is fundamentally still an
isophotal magnitude and therefore underestimates the integrated light
of galaxies \citep{dalcanton98b}.

Metric magnitudes appear to offer a physically motivated solution to
the shortcomings of these two techniques. The most common metric magnitude
is measured within the radius introduced by \citet{petrosian76}. The Petrosian
magnitude is defined to be the flux interior to radius $r_p$, where
the surface brightness at $r_p$ is equal to some fraction of the average
surface brightness interior to $r_p$. The great attraction of
using a Petrosian magnitude is that it is independent of the spacetime 
geometry by 
virtue of being the ratio of two surface brightnesses. $r_p$
therefore corresponds to the same physical scale in two galaxies with the
same surface-brightness profile at different redshifts. The corresponding
integrated brightness within that radius will correspond to the same
fraction of the total light (any systematic underestimate will be identical
for the two galaxies). However, the fraction of the light missed by a
Petrosian magnitude for two galaxies at different redshifts will only be the
same if two conditions hold: the surface
brightness profiles of the galaxies are the same and neither of the galaxies
are small enough that they are significantly affected by the PSF. Thus while
the Petrosian magnitude is ideal for cosmological tests \citep{sandage90,
petrosian98}, it may not offer the ideal choice for measuring the integrated
light of galaxies with a diverse range of surface brightness profiles and
sizes comparable to the PSF of the observations.

The main goal of the magnitude measurements in this paper is to measure the
number of galaxies per unit magnitude. These measurements,
in conjunction with theoretical predictions, can then be used to
constrain models of galaxy formation and evolution. The output of the
theoretical models to which observed number counts are compared is the sum of
galaxies in a given apparent magnitude bin. To compare observational data to
these theoretical predictions therefore requires an unbiased measurement of the
number of galaxies at a given integrated brightness, which in turn requires a
good means of estimating the integrated brightness of individual galaxies.

I evaluated different techniques for measuring the
integrated brightness of the galaxies in the sample including aperture
magnitudes measured with PHOT in IRAF, TOTAL magnitudes measured with
FOCAS, and ``BEST'' magnitudes measured by SExtractor. The SExtractor 
BEST magnitude is defined to be either a Kron magnitude \citep{kron80}, which
is measured in an elliptical aperture whose size is determined by the
object's profile, or an isophotal magnitude in very crowded regions.
To evaluate the accuracy of these photometry techniques I used ARTDATA in
IRAF (as described in \S \ref{sec:det}) to add artificial galaxies to
the images. The metric to evaluate the accuracy of each of these techniques
is the size of the difference between the input magnitude and the
measured magnitude as a function of $r_h$ for exponential and $\dv$
profiles. I performed these tests for different images to sample the range
of seeing in the data and at a range in apparent magnitude from the
50\% completeness limit for stellar profiles to four magnitudes
brighter. This magnitude range includes more than 80\% of the sources
in the sample. The investigation showed that IRAF PHOT aperture magnitudes, 
FOCAS TOTAL magnitudes, and SExtractor BEST magnitudes all underestimate the
integrated brightness of galaxies. The FOCAS and SExtractor magnitudes
were taken directly from the output of these two packages, while I added a
stellar-profile aperture correction, based on measurements of several stellar
objects in the field, to the aperture magnitudes measured by IRAF PHOT.

\begin{figure*}[t]
\centerline{
\epsfxsize=3.5truein\epsfbox[65 165 550 730]{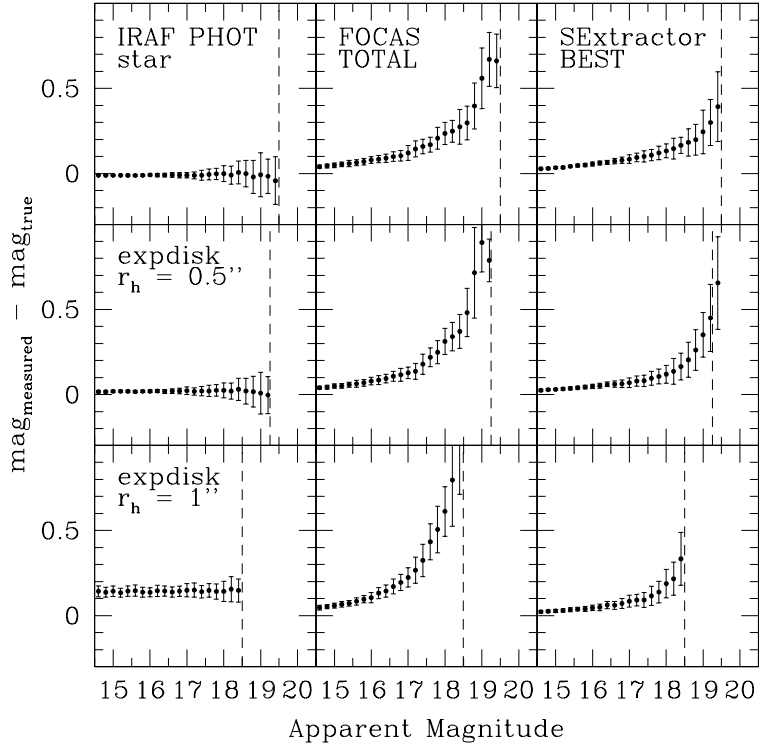}}
\caption{
\footnotesize
Reliability of the photometric methods as a function of image size and
integrated brightness. The left panels show the difference between the 
measured and true magnitudes for aperture photometry with IRAF PHOT. The 
top panel shows the results for stellar objects, the middle and bottom panels 
for exponential disks with $r_h = 0.5''$ and $r_h = 1''$, respectively. The 
middle panels show the same experiment with FOCAS TOTAL magnitudes and the 
right panels with SExtractor BEST magnitudes. The dashed, vertical line is 
the 90\% detection limit for the objects in each panel. As in 
Figure~\ref{fig:det}, the sky noise and magnitude zeropoint were taken from 
one of the $K$ band frames; however, these results are only dependent on the 
apparent magnitude relative to the magnitude of the 90\% detection limit. 
} \label{fig:phot}
\end{figure*}

Figure~\ref{fig:phot} shows the result of one of these experiments
for an image with 1.25$''$ FWHM. The measured magnitude is either the
IRAF PHOT aperture magnitude ({\it left panels}),
FOCAS TOTAL magnitude ({\it middle panels}), or
SExtractor BEST magnitude ({\it right panels}) and the true magnitude
is the input magnitude assigned to the artificial object. 
As in Figure~\ref{fig:det}, the sky noise and apparent magnitude for this 
simulation are based on one of the $K$ band frames; however, the reliablility 
of these different photometric techniques relative to the completeness limit 
are independent of the filter. 
This figure
shows the results for stellar profiles ({\it top panels}), $r_h = 0.5''$
exponential disks ({\it middle panels}), and $r_h = 1''$ ({\it bottom
panels}). The data points are the mean offset between the measured and
true magnitudes from $\sim 200$ artificial objects per apparent magnitude
and the errorbars are the $1\sigma$ dispersion in these measurements at
that magnitude.  At a fixed input galaxy brightness more light is missed
in galaxies with larger $r_h$, while at fixed $r_h$, more light is missed
in an $\dv$ profile than in an exponential profile.
For fainter apparent magnitudes, the size of the brightness underestimate
increases for the FOCAS TOTAL and SExtractor BEST magnitudes at fixed $r_h$, 
but remained constant for the aperture magnitude.  The reason for the 
systematic increase in the underestimate for the FOCAS TOTAL magnitudes is 
that the isophotal 
area of a galaxy of fixed size decreases as the integrated brightness decreases.
Though the FOCAS TOTAL magnitude doubles the isophotal area, this still leads to
an underestimate of the true size of the galaxy for the photometric
measurement. The SExtractor BEST magnitude suffers from the same systematic 
increase in error as the size of the elliptical aperture is set by a fit to the
observed galaxy profile. This offset is also due to the fact that SExtractor 
BEST
magnitudes assume the PSF is Gaussian when computing the elliptical aperture
for photometry and a Gaussian profile is more centrally concentrated than a
MOFFAT profile \citep{dalcanton98b}. At a higher galaxy surface density than 
probed in this survey, the SExtractor BEST magnitudes will be isophotal 
magnitudes, rather than Kron magnitudes, and the brightness underestimate may 
be more severe.  For fainter galaxies at fixed size and profile
shape, a smaller fraction will be above the noise and the tendency is for a
detected object to appear more compact, which will further shrink the aperture
size used in the photometry. While I have not evaluated Petrosian
magnitudes here, \citet{dalcanton98a} included them in her study of
systematic biases in measuring the integrated light of galaxies for computing
the luminosity function. She found that Petrosian magnitudes underestimate
the true brightness of galaxies to a lesser extent than aperture or isophotal
magnitudes \citep[see also][]{dalcanton98b}, although they too are subject
to large errors for objects near the detection limit. This is because the
surface brightness profile becomes increasingly noisy. In addition, as stated
above, the Petrosian radius is difficult to determine for marginally resolved
objects.

\citet{saracco99} also compare aperture magnitudes to isophotal magnitudes and
SExtractor AUTO magnitudes, which are similar to SExtractor BEST magnitudes.
Their figure~1 shows a difference of several tenths of a magnitude between
photometric measurements with these three techniques of the galaxies in
their fields. These differences increase to on order one magnitude for the
last two magnitudes below their S/N = 5 detection limit for compact objects.
They find that both isophotal and SExtractor AUTO magnitudes underestimate the 
flux of faint objects and therefore chose to use aperture magnitudes. 
\citet{vaisanen00} have also evaluated the relative merits of aperture,
isophotal, and SExtractor BEST magnitudes. They concluded that 
SExtractor BEST magnitudes do not
underestimate the integrated brightness of galaxies near the detection limit,
in contrast with the simulations presented here and the results of
\citet{dalcanton98b}.

My analysis has shown that only aperture magnitudes avoid errors in the
photometry that are a function of integrated brightness, and I have
chosen to use them. The brightness underestimate of aperture photometry
will still depend on the intrinsic size and profile shape, but avoids the
additional systematic uncertainty due to the dependence on integrated
brightness inherent in SExtractor BEST and FOCAS TOTAL magnitudes.  
Another consideration in
the decision to use aperture magnitudes is the unique nature of
these data. I have 21 fields observed in three filters, each with a different
PSF shape.  Aperture photometry is the best way to account for the
heterogeneous nature of the dataset.  The size of the error in the 
SExtractor BEST and FOCAS TOTAL magnitudes will 
vary from field to field with the seeing.
While aperture magnitudes will suffer from the same problem, they can be
relatively easily scaled to take seeing variations into account. In addition,
the aperture magnitudes can be easily measured through a sufficiently large
aperture to minimize the size of the galaxy aperture correction and therefore
the associated uncertainty.

Given the range of galaxy sizes and typical seeing, aperture magnitudes with
a radius 1.5 times the FWHM of the seeing are a good compromise between
minimizing the size of the galaxy aperture correction and maximizing the
galaxy signal in the aperture. At $z = 0.5$, 1$''$ corresponds to a
physical size of 4.3 $h^{-1}$ kpc (for $\om = 0.3$, $\ol = 0.7$) and therefore
these typical aperture sizes of 1.5$''$ -- 2.5$''$ include most of the
galaxies' light.
The galaxy aperture correction is the difference between the true
integrated brightness of the galaxy and the measured aperture magnitude
plus the stellar aperture correction. The stellar aperture correction
only depends on the shape of the PSF and, because the aperture is scaled to the
PSF size, this correction was always measured to be $\sim 0.1$ mag.
The galaxy aperture correction depends on the typical galaxy size and was
measured with the same code developed to measure the detection efficiency
for objects of different size, surface brightness profile, and
integrated brightness. The distribution of size and surface brightness
profile as a function of integrated brightness have not been completely
determined in the NIR. However, a range of studies of
galaxy morphology at visible \citep{smail95,roche97,roche98}
and NIR \citep{yan98,bershady98,saracco99} wavelengths,
combined with color information \citep{thompson99} shows that
most of the faint galaxies in this survey will have exponential
profiles with typical half-light radii of $0.5''$ --  $0.75''$. 
At the magnitude limits of this survey the average galaxy size is 
$r_h \sim 0.6''$ in these three filters. 
Any bulge
component for these galaxies, though more prominent at NIR
wavelengths, will fall inside the aperture. For this range of sizes of
galaxies, I computed the galaxy aperture correction for each frame as
a function of integrated brightness. The size of this correction is
always in the range 0.1 -- 0.2 mag. The correction is generally smaller
for fields with poorer seeing as the aperture size is larger in these
fields, although the detection limit for a given galaxy size and profile
shape is lower for these fields as well.

\section{Star/Galaxy Separation} \label{sec:stars}

The calculation of number counts of galaxies at faint magnitudes must deal
with stellar contamination, which affects faint galaxy surveys in two
ways. First, the typical galaxy size decreases at fainter apparent magnitudes
and therefore galaxies are
progressively more likely to be morphologically indistinguishable from stars.
Second, stars (and compact galaxies) are the easiest objects to detect and
accurately photometer near the survey limits. This is illustrated in
Figure~\ref{fig:det}, which shows that the limiting magnitude at fixed
detection efficiency is on order one magnitude brighter for exponential
disks or $\dv$ profiles with $r_h = 1''$ than for stars. Multicolor separation 
is one way to remove stellar contamination from faint galaxy samples. 
This process is most effective with small photometric errors, although it can 
be effective with larger errors with a longer wavelength baseline 
\citep[e.g. visible--NIR colors,][]{gardner95}.  The removal of stellar 
contamination from this survey is particularly important as stars contribute 
on order 20 -- 30\% to even the faintest magnitude bins of this survey. 
In even deeper surveys stellar contamination is less of a problem as 
galaxies begin to significantly outnumber stars.

The stellar content of the DMS fields has already been cataloged as part
of the search for quasars \citep{hall96a,osmer98b}. The selection criterion for
identifying stellar objects was that they must be indistinguishable from the
PSF in at least three of the six CCD filters \citep{hall96a}. I used the
stellar catalog from \citet{osmer98b} to identify and filter out stars from
the NIR catalogs.
Given that the CCD stellar catalog extends several magnitudes fainter than
this survey, all but the very reddest stars should be included in the
stellar catalog. In contrast, the CCD stellar catalog includes some
contamination from compact galaxies near its magnitude limit. From the
stellar model counts presented in \citet{minezaki98a} and galaxy
counts in the literature, I expect stars comprise approximately 10 -- 20 \%
of all objects at the limiting magnitude of this survey. Any missed fraction
will therefore be a minor contamination to the galaxy number counts.

\section{Number Counts} \label{sec:num}

The calculation of NIR galaxy number counts in the DMS is complicated
by the variation in detection efficiency from field to field. To combine
all of the fields I have used a technique analogous to the one outlined in
the appendix of \citet{bershady98}. However, the same variations in seeing
that contribute to the different field sensitivities make it difficult to
reliably bin these objects into different size classes. I have therefore not
separated the counts into different sizes (and profile types), effectively
collapsing the galaxy distribution from a multivariate distribution (size,
profile shape, and integrated brightness) to a function of integrated
brightness alone. As most number--magnitude models for galaxies only
predict surface density vs.\ integrated brightness \citep[though
see][]{im95}, this is the most useful format to compare the
data with theoretical predictions.

\begin{figure*}[t]
\centerline{
\epsfxsize=3.5truein\epsfbox[65 165 550 730]{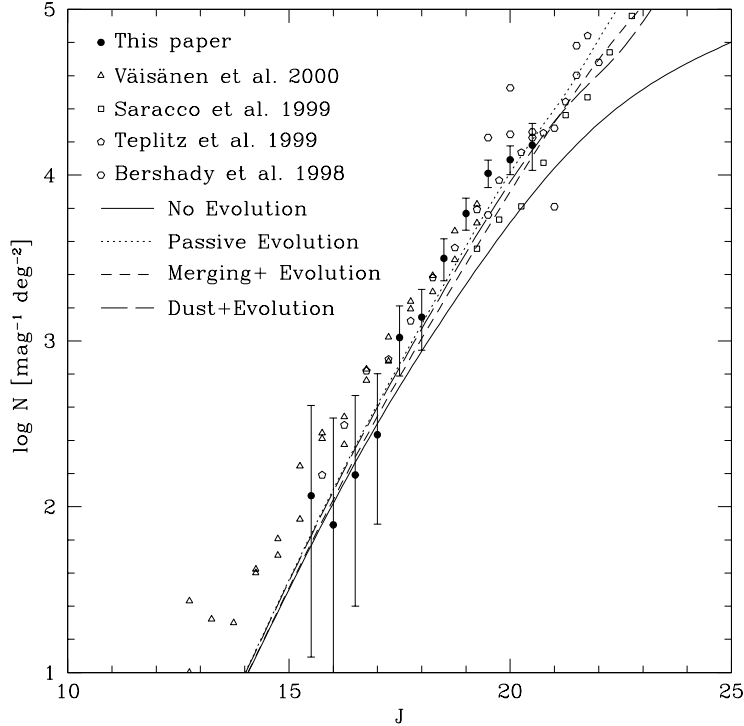}}
\caption{
\footnotesize
Differential galaxy number counts in the near-infrared $J$ band. The raw
number counts have been converted into units of mag$^{-1}$ deg$^{-2}$
({\it filled circles}) and the errorbars are 1$\sigma$ confidence limits.
These data are listed in Table~\ref{tbl:numj}. Also shown are number counts
from the literature and four models from \citet{gardner98}.
} \label{fig:numj}
\end{figure*}

\begin{figure*}[t]
\centerline{
\epsfxsize=3.5truein\epsfbox[65 165 550 730]{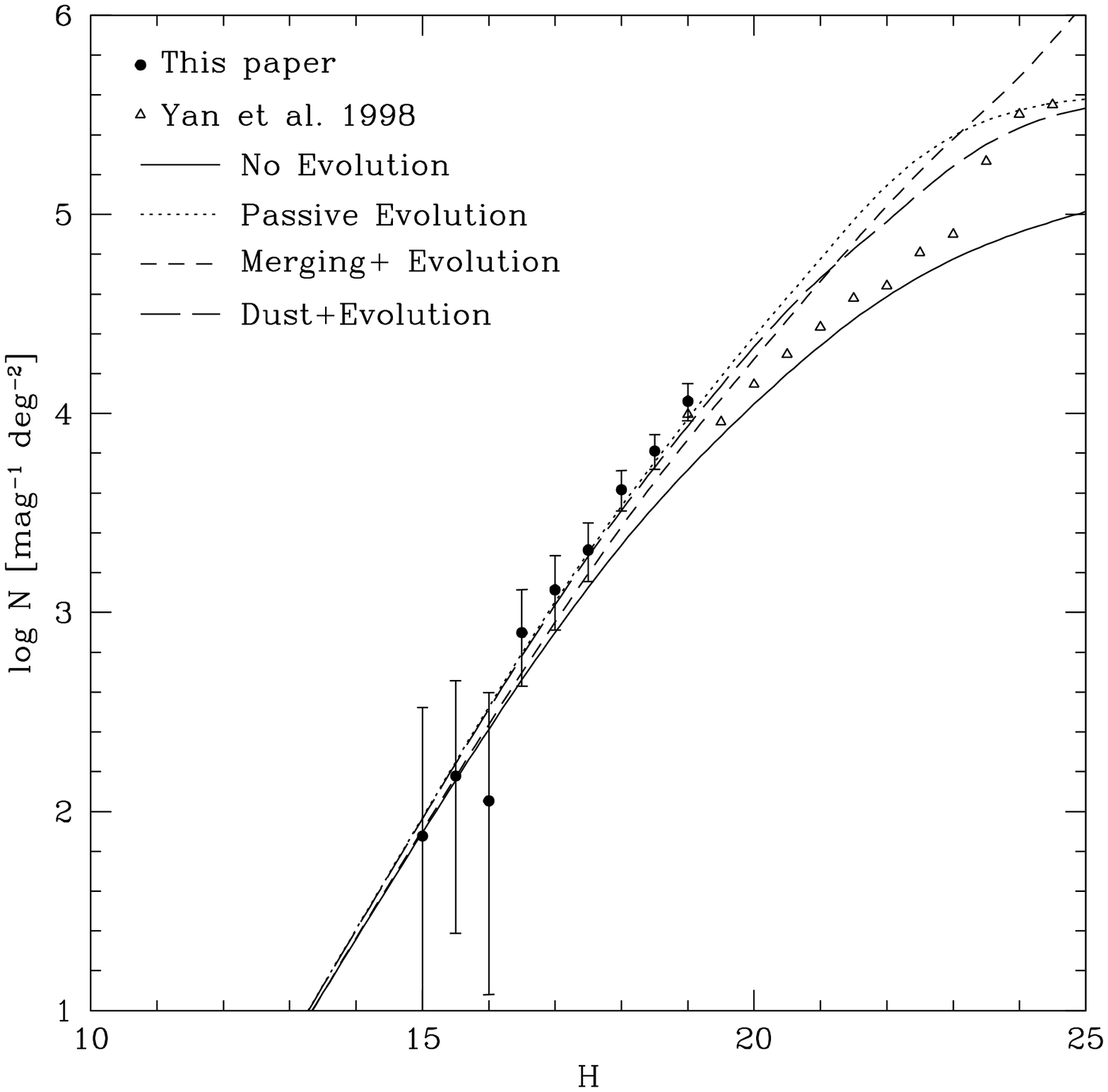}}
\caption{
\footnotesize
Differential galaxy number counts as in Figure~\ref{fig:numj} but for
the near-infrared $H$ band.
} \label{fig:numh}
\end{figure*}

\begin{figure*}[t]
\centerline{
\epsfxsize=3.5truein\epsfbox[65 165 550 730]{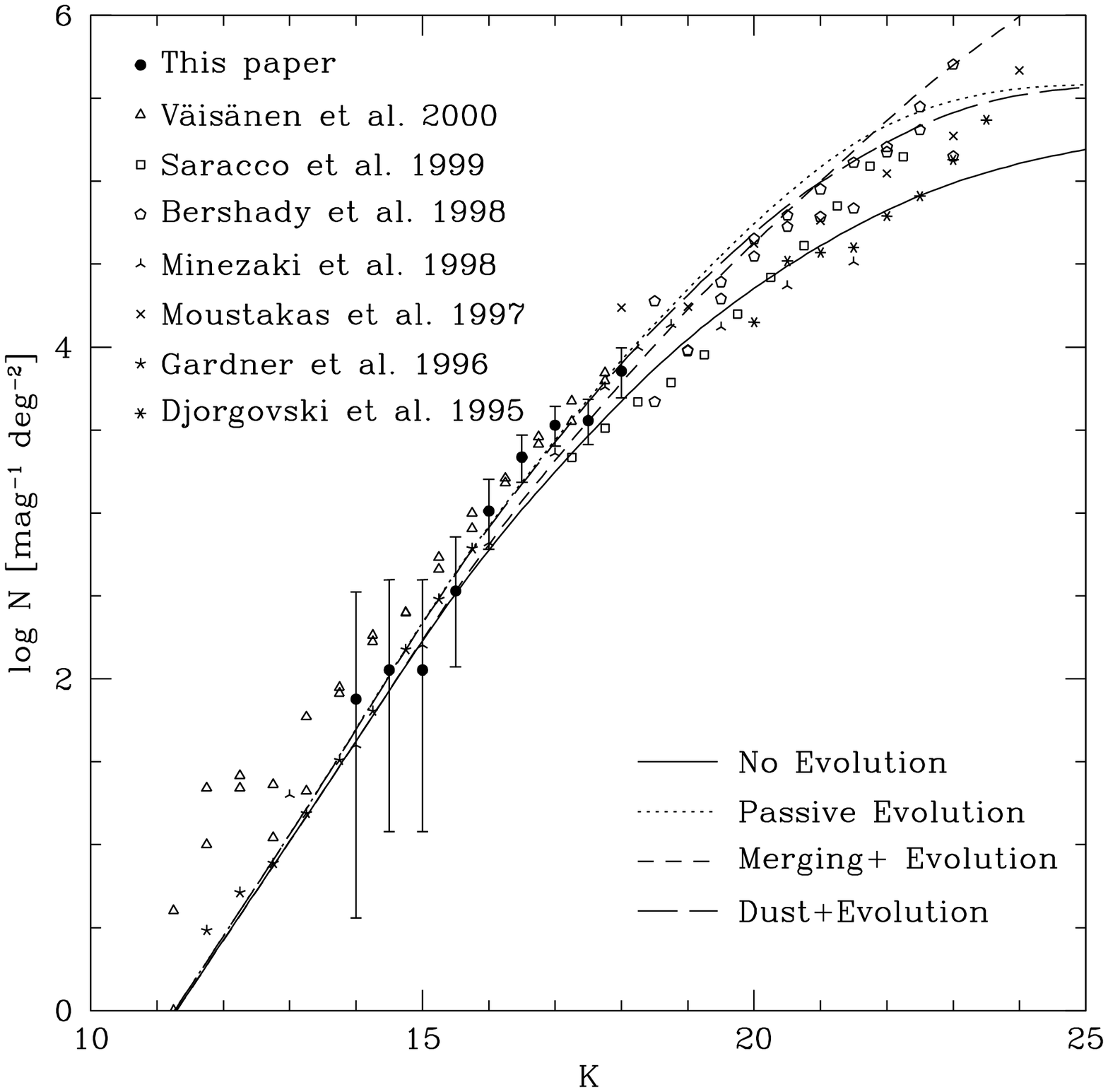}}
\caption{
\footnotesize
Differential galaxy number counts as in Figure~\ref{fig:numj} but for
the near-infrared $K$ band.
} \label{fig:numk}
\end{figure*}

For each apparent magnitude bin I used the typical galaxy size and galaxy
aperture correction discussed in \S \ref{sec:phot} to correct the measured
brightness of each galaxy. All of the galaxies in each field were then
placed into 0.5 mag bins in apparent magnitude.
The detection efficiency for each bin is the average of the detection
efficiency vs.\ magnitude measured in 0.1 mag increments and weighted by
the expected slope of the number--magnitude relation $\alpha \sim 0.5$,
where ${\rm log} N \propto \alpha \, m$. In practice, the detection efficiency
drops off significantly faster than the number--magnitude relation increases
and so the detection efficiency of a magnitude bin is relatively insensitive
to $\alpha$. If the detection efficiency of the faint end of a bin was
less than 50\%, this bin was not included in the calculation of the
number--magnitude relation.
The detection efficiencies described in \S \ref{sec:det} multiplied
by the field size yields the effective area of each field. I then
summed the total differential counts and effective area of each field per
0.5 magnitude bin to compute the differential number counts per unit magnitude
per square degree. The number counts for $J$, $H$, and $K$ are shown in
Figures~\ref{fig:numj}, \ref{fig:numh}, and \ref{fig:numk}, respectively.
The raw and corrected counts, along with the effective area, are listed in
Tables~\ref{tbl:numj}, \ref{tbl:numh}, and \ref{tbl:numk}.
The raw counts are the total counts in 0.5 mag bins with only the
stellar ($\sim 0.1$ mag) aperture correction applied to the photometry.
The corrected counts are the counts per 0.5 mag bin after application
of the galaxy aperture correction.
The errorbars on the number counts are $1\sigma$ upper and lower
confidence intervals calculated using the formulae in \citet{gehrels86}.

The $J$ number counts shown in Figure~\ref{fig:numj} agree well with the
number counts of \citet{teplitz99}, \citet{vaisanen00} and the 
(small+large) counts of
\citet{bershady98}, though they are somewhat higher than the \citet{saracco99}
counts. The \citet{vaisanen00} survey overlaps the most with this survey, 
although they use SExtractor BEST magnitudes rather than aperture 
magnitudes. Their survey 
covers a substantially larger area than this one, though its faint limit is 
nearly one magnitude brighter. \citet{bershady98} and \citet{saracco99} 
extend much fainter than this survey and their brightest counts only 
overlap with the faintest magnitudes of this survey and in this regime 
they have approximately five to ten times fewer galaxies due to their 
smaller area. The \citet{saracco99} surface density is probably lower in 
this range due to small number statistics as they also use the 
\citet{persson98} standards, aperture magnitudes, and the agreement appears 
to improve at fainter magnitudes where they have a larger sample. 
All of the $J$ counts in the range 16 -- 20.5 mag have essentially the
same slope as these data, $\alpha \sim 0.54$. 

The slope of the $H$ band counts in Figure~\ref{fig:numh} over the 16.5 -- 19 
mag range is $\alpha = 0.47$. These counts overlap with the brightest 
magnitude bin from \citet{yan98}, but the surface density vs.\ magnitude 
relation shows an apparent ``break'' at $H = 19$ -- 20 mag. 
The \citet{yan98} measurements are based on NICMOS parallel data with 
Camera 3 over a total area of 8.7 arcmin$^2$ and therefore they have 
relatively few galaxies in the region of overlap. However, the additional 
NICMOS parallel data studied by \citet{teplitz98}, which includes 
images from Camera 2 and extends several magnitudes brighter, also 
shows an apparent break at $H = 19$ -- 20 mag. \citet{teplitz98} transformed 
several deep $K$ galaxy counts studies to the NICMOS F160W filter 
and these transformed $K$ counts appear to match the break in the 
$H$ galaxy number--magnitude relation. This break may therefore be the change
in slope seen in $K$ band counts at $K \sim 20$ mag. 

Most published NIR number counts to date have been obtained in the
$K$ band \citep{gardner93,cowie94,glazebrook94,soifer94,djorgovski95,mcleod95,
moustakas97,minezaki98b,szokoly98,bershady98,saracco99,vaisanen00,
mccracken00}.  
\citet{gardner96} and \citet{vaisanen00} have the largest magnitude range 
of overlap with the DMS counts shown in Figure~\ref{fig:numk} and the 
number--magitude relations from these studies are in good agreement. 
The slope of the $K$ counts from 14 -- 18 mag is $\alpha = 0.54$. 
At fainter magnitudes there is a great deal of dispersion in the 
surface density measurements, though given the small areas of these surveys 
they are generally consistent with one another. As noted by several 
authors \citep[e.g.][]{bershady98,saracco99}, this dispersion may also be 
due to different measurement techniques, variations in magnitude systems, and 
filters, in addition to small number statistics and cosmic variance. 
The most significant differences are likely due to the method of magnitude 
measurement, completeness corrections, and at least at bright magnitudes, 
star--galaxy separation. Of the faint counts presented in 
Figure~\ref{fig:numk}, \citet{saracco99} and \citet{moustakas97} apply a 
uniform galaxy aperture correction for all objects, \citet{bershady98} compute 
completeness and a galaxy aperture correction as a function of object size, 
\citet{minezaki98b} use FOCAS TOTAL magnitudes, and \citet{djorgovski95} only use a 
stellar aperture correction. 

I computed the four models shown in Figures~\ref{fig:numj} -- \ref{fig:numk}
with {\it ncmod}, a publically available code for computing galaxy number
counts and colors by \citet{gardner98}. The models correspond to no evolution
({\it solid lines}), passive evolution ({\it dotted lines}), merging plus
passive evolution ({\it short-dashed lines}), and dust plus passive
evolution ({\it long-dashed lines}).
For all of the models I have assumed $\om = 0.3$ and
$\ol = 0.7$, in agreement with results on Type Ia
supernovae \citep{riess98,perlmutter99}, the Lyman-$\alpha$ forest,
COBE-DMR \citep{phillips00}, as well as number counts in the {\it Hubble
Deep Field} \citep{totani00}. I chose to fix the cosmological parameters
as the uncertainties in $\om$ and $\ol$ are less than the 
uncertainties in models of galaxy evolution.

All of these models are based on the $K$ luminosity function and
mix of galaxy types from \citet{gardner97} and galaxy spectral energy
distributions from GISSEL96 \citep{bruzual93}.
To compute these models I also used the filter transmission curves for the
TIFKAM filters plus a measurement of the atmospheric transmission.
The no evolution
model is a pure $k$-correction model. In the passive evolution model, all
galaxies form at $z = 15$ except for the latest galaxy type, which have
constant star formation and are always 1 Gyr old. The merging model has
the same galaxy evolution as the passive evolution mode, but it also
includes number evolution in the form $\phi^* \propto (1 + z)^\eta$ with
$\eta = 1.5$ and conservation of luminosity density. This model is
based on the parameterization of \citet{rocca90} and it produces similar
results to the \citet{broadhurst92} merging model. This value of $\eta$ is
higher than the constraint $\eta \leq 1$ derived by \citet{totani00} in their
$\Lambda$ model for the HDF, but is illustrative of the effect of adding
merging to passive evolution. Finally, the dust model also contains the same
galaxy evolution parameterization as the passive evolution model, but with
the addition of dust in the plane of the galaxies. This model, based on the
work of \citet{bruzual88} and \citet{wang91} was introduced to number count
models to correct for their tendency to overproduce the $UV$ flux in
galaxies, though it has a minimal effect on the NIR number--magnitude
relation.

The $J$, $H$, and $K$ observations presented here all agree well with the
passive evolution model. They are less consistent with the
number density evolution we assumed in the merging model, and clearly
have a higher surface density of objects than the no evolution prediction.
The observed $J$ counts are all higher than the models, but this
may be due to the extrapolation of the $K$ galaxy luminosity function to
the $J$ band. As noted above, the {\it HST} counts for $H > 20$ mag
\citep{teplitz98,yan98} have systematically lower surface density. This
difference does not appear to be due to differences in the transmission
profiles of the two filters. When I calculated the model galaxy counts with
the F160W transmission profile, rather than the $H$ filter, the F160W galaxy
count models were offset by less than 0.1 mag fainter than the ground-based
$H$ model counts, whereas an offset of  0.3 -- 0.4 mag is
needed to explain the difference in surface density.

\section{Mean Colors} \label{sec:colors}

\begin{figure*}[t]
\centerline{
\epsfxsize=6truein\epsfbox[0 280 590 600]{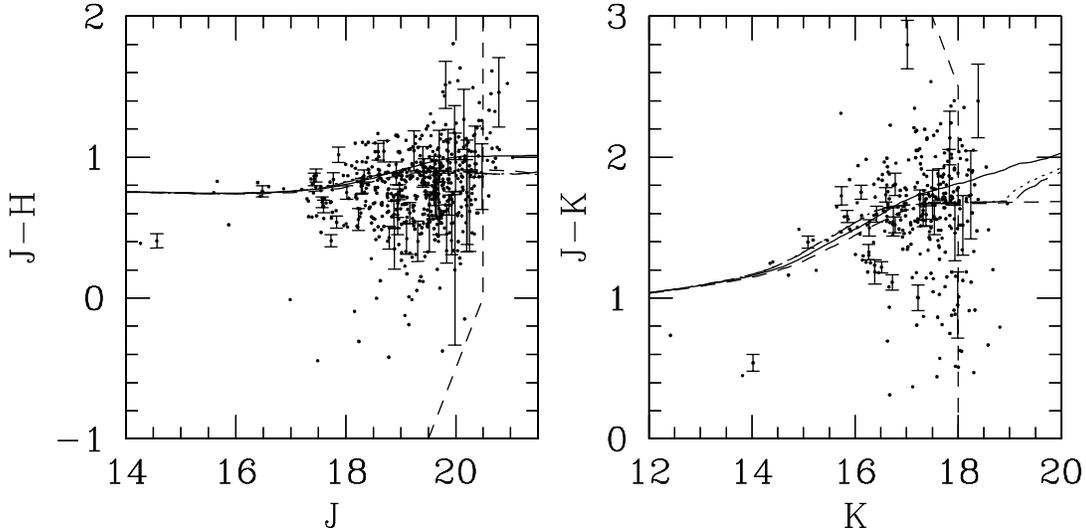}}
\caption{
\footnotesize
Mean near-infrared colors vs.\ magnitude for the galaxies in this sample.
The data are represented by points with errorbars. Only every tenth point 
has an associated errorbar for clarity. The curves are the mean
colors vs.\ magnitude for the four models shown in Figures~\ref{fig:numj} --
\ref{fig:numk} and discussed in \S\ref{sec:num}. The straight,
dashed lines are the average 50\% limiting magnitudes and colors for an 
exponential disk with $r_h = 0.75''$.
The scatter of the data points about the model lines is primarily due to
variance in the typical galaxy colors, rather than photometric uncertainty.
} \label{fig:col}
\end{figure*}

Color as a function of apparent magnitude can provide an additional constraint
on galaxy evolution models, particularly when severals colors are available.
I have used the \citet{gardner98} model to compute the median $J-H$ color
vs.\ $J$ magnitude and the median $J-K$ color vs.\ $K$ magnitude for the
four models discussed above. Figure~\ref{fig:col} shows these two
color--magnitude relations for all of the galaxies detected in these filters.
For the magnitude range of this survey, all of these models predict similar
median colors and they agree well with the data and $J-K$ colors reported
by \citet{bershady98} and \citet{saracco99} in the region of overlap.

To these apparent magnitude limits, NIR colors do not provide a strong
constraint on galaxy evolution as most galaxies are low redshift
($z < 1$) objects. In this redshift range, the $JHK$ filters all sample the
old stellar population, which has a relatively flat spectral energy
distribution independent of galaxy type. The overlap of the four models and
the lack of change in the median color with apparent magnitude reflects the
similarity of these galaxies at NIR wavelengths.  \citet{bershady98} and
\citet{saracco99} also report $J-K$ colors vs.\ $K$ and these results agree
well with their colors. 

\section{Summary} \label{sec:sum}

I have presented NIR $JHK$ galaxy counts and colors and described
in detail the detection efficiencies and selection effects that affect this
survey. I have studied several popular methods for measuring the integrated
brightness of galaxies as a function of apparent magnitude and found that
only aperture magnitudes miss a consistent fraction of a galaxy's light
as a function of apparent magnitude for a range of galaxy sizes and profile 
shapes. This survey is the largest area observed to these depths in all 
three NIR filters that has been published to date. The $J$, $H$, and $K$ 
number--magnitude relations and colors are in good agreement with a simple 
model of passive galaxy evolution with at most a small amount 
($\eta \leq 1.5$) of merging in an $\om = 0.3$, $\ol = 0.7$ cosmology.

Current model predictions for the number--magnitude relation at 
NIR wavelengths are very similar to one another and larger surveys 
at the faintest magnitudes are needed to discriminate between different 
models with this diagnostic tool. The models shown in Figures~\ref{fig:numj} 
-- \ref{fig:numk} suggest that number counts to $H \sim 24$ mag or $K \sim 23$
mag are necessary to distinguish between the passive evolution and 
merging model. High angular resolution imaging to measure the mix of galaxy 
sizes and morphological types as a function of brightness, as well as color 
information, could prove to be a more efficient means to discriminate between 
galaxy formation and evolution models with deep, NIR imaging surveys.

\acknowledgements

I would like to thank Darren DePoy, Patrick Osmer, and David Weinberg for 
helpful discussions and suggestions as well as the referee, 
Jonathan Gardner, for helpful comments that have improved this presentation. 
I am grateful to Christopher J. Burke and Adam Steed for assistance with 
some of the observations, and to Mike Lannon for supplying the TIFKAM filter 
curves in digital form. 
I would also like to thank the MDM staff for their support and the OSU
TAC for a generous allocation of telescope time. 
I was supported in part by a Presidential Fellowship from Ohio State 
University and received additional travel and other support from a PEGS grant 
and the Department of Astronomy at Ohio State University. 
TIFKAM was funded by the Ohio State University, the MDM consortium, MIT, and 
NSF grant AST-9605012. NOAO and USNO paid for the development of the ALADDIN 
arrays and contributed the array currently in use in TIFKAM. 

{}

\begin{center}
\begin{deluxetable}{lclllccc}
\tablenum{1}
\tablewidth{0pt}
\tablecaption{Log of Observations \label{tbl:log} }
\tablehead {
  \colhead{Field} &
  \colhead{Area} &
  \colhead{Date$_J$} &
  \colhead{Date$_H$} &
  \colhead{Date$_K$} & 
  \colhead{FWHM$_J$} &
  \colhead{FWHM$_H$} &
  \colhead{FWHM$_K$} \\
}
\startdata
CF1	& 4.52\tablenotemark{a}	& 990928 & 990929 & 990929 & 1.4 & 1.6 & 1.5 \\
CF3 	& 6.94	& 971014 & 971016 & 971014 & 1.8 & 1.2 & 1.8 \\
21WC	& 7.74	& 971015 & 990926 & 990926 & 1.4 & 1.4 & 1.3 \\
22EC	& 7.46 	& 971016 & 971017 & 971016 & 1.2 & 1.3 & 1.2 \\
10EC	& 9.49 	& 980411 & 980411 & 980410 & 1.5 & 1.4 & 1.2 \\
14NC	& 7.55	& 990515 & 980411 & 980410 & 1.2 & 1.5 & 1.2 \\
17NC	& 8.71	& 980411 & 980415\tablenotemark{b} & 980410 & 1.5 & 1.1 & 1.2 \\
14SC	& 9.17	& 980414\tablenotemark{c} & 980414\tablenotemark{c} & 980414\tablenotemark{c} & 1.3 & 1.2 & 1.6 \\
22WC	& 9.23	& 981003 & 981003 & 981003 & 1.6 & 1.6 & 1.2 \\
01EC60S & 8.81 	& 981003 & 981002 & 981002 & 1.3 & 1.2 & 1.3 \\
01WC	& 9.94	& 981003 & 981002 & 981002 & 1.3 & 1.3 & 1.5 \\
21EC	& 9.31	& 981001 & 981001 & 981002 & 1.4 & 1.2 & 1.5 \\
01WC150W & 9.86	& 981003 & 981003 & 981003 & 1.4 & 1.5 & 1.5 \\
21WC150W & 9.30	& 981004 & 981004 & 981004 & 1.7 & 1.5 & 1.4 \\
14NC150W & 9.04 & 990516 & 990516 & 990516 & 1.0 & 1.0 & 1.0 \\
14NC150E & 9.36 & 990517 & 990517 & 990517 & 1.6 & 1.0 & 1.2 \\
14NC300W & 9.90 & 990519 & 990519 & 990519 & 1.2 & 1.1 & 1.2 \\
14NC300E & 9.54 & 990520 & 990520 & 990520 & 0.9 & 0.9 & 1.0 \\
17SC	& 9.70 	& 990515 & 990926 & 990927 & 1.1 & 1.1 & 1.1 \\
21WC150E & 9.80 & 990928 & 990928 & 990928 & 1.3 & 1.3 & 1.3 \\
22EC150W & 10.19 & 990929 & 990929 & 990929 & 1.7 & 1.8 & 1.9 \\
\enddata
\tablenotetext{a}{Area that does not overlap 01WC150W}
\tablenotetext{b}{Calibration obtained 981003}
\tablenotetext{c}{Calibration obtained 990515}
\tablecomments{Log of the observations of the 21 subfields of the DMS.  
Column~1 lists the field identifier, which indicates either the center of one 
of the fields listed in \citet{hall96a} or the offset from the center of one 
of these fields. For example, 21WC refers to the center of Field 21W and 
01EC60S refers to a field 60$''$ south of Field 01E. 
The two exceptions to this naming convention are CF1 and CF3, which are defined in \citet{hall98}. 
Column~2 has the area in square arcminutes of each subfield, while columns 
3 -- 5 the UT date of the observations in $J$, $H$, and $K$, respectively. 
Columns 6 -- 8 list the FWHM in arcseconds of the PSF in the final, combined 
frame for $J$, $H$, and $K$. 
}
\end{deluxetable}
\end{center}

\begin{center}
\begin{deluxetable}{lcccccccccccc}
\tablenum{2}
\tablewidth{0pt}
\tablecaption{Photometric Solutions \label{tbl:phot} }
\tablehead {
  \colhead{Date} &
  \colhead{$j_0$} &
  \colhead{$j_1$} &
  \colhead{$j_2$} &
  \colhead{$\sigma(J)$} &
  \colhead{$h_0$} &
  \colhead{$h_1$} &
  \colhead{$h_2$} &
  \colhead{$\sigma(H)$} &
  \colhead{$k_0$} &
  \colhead{$k_1$} &
  \colhead{$k_2$} &
  \colhead{$\sigma(K)$} \\
}
\startdata
971011  &  22.204  &  -0.040  &  0.160  &  0.046  &  21.961  &  0.000  &  0.060  &  0.074  &  21.245  &  -0.060  &  0.000  &  0.127  \\
971014  &  22.303  &  -0.047  &  0.160  &  0.036  &  22.040  &  -0.006  &  0.060  &  0.048  &  21.290  &  -0.060  &  0.000  &  0.048  \\
971015  &  22.351  &  -0.038  &  0.160  &  0.048  &  22.076  &  -0.021  &  0.060  &  0.029  &  21.354  &  -0.053  &  0.000  &  0.077  \\
971016  &  22.346  &  -0.040  &  0.160  &  0.036  &  22.063  &  -0.040  &  0.060  &  0.034  &  21.324  &  -0.060  &  0.000  &  0.064  \\
971017  &  22.330  &  -0.040  &  0.160  &  0.031  &  22.070  &  0.000  &  0.060  &  0.035  &  21.320  &  -0.060  &  0.000  &  0.039  \\
980410  &  22.425  &  -0.113  &  0.060  &  0.034  &  22.033  &  -0.058  &  0.060  &  0.036  &  21.316  &  -0.050  &  0.000  &  0.035  \\
980411  &  22.409  &  -0.206  &  0.060  &  0.022  &  22.047  &  -0.158  &  0.060  &  0.028  &  21.311  &  -0.134  &  0.000  &  0.048  \\
981001  &  22.293  &  -0.060  &  0.060  &  0.032  &  22.075  &  -0.037  &  0.000  &  0.021  &  21.372  &  -0.065  &  0.000  &  0.032  \\
981002  &  22.336  &  -0.070  &  0.060  &  0.019  &  22.147  &  0.000  &  0.000  &  0.035  &  21.485  &  -0.052  &  0.000  &  0.017  \\
981003  &  22.355  &  -0.081  &  0.060  &  0.032  &  22.164  &  -0.026  &  0.000  &  0.020  &  21.485  &  -0.091  &  0.000  &  0.033  \\
981004  &  22.332  &  -0.102  &  0.060  &  0.021  &  22.165  &  -0.058  &  0.000  &  0.022  &  21.470  &  -0.113  &  0.000  &  0.024  \\
990515  &  22.377  &  -0.090  &  0.060  &  0.040  &  22.182  &  0.000  &  0.000  &  0.017  &  21.494  &  -0.050  &  0.000  &  0.029  \\
990516  &  22.412  &  -0.090  &  0.060  &  0.032  &  22.181  &  -0.008  &  0.000  &  0.025  &  21.505  &  -0.046  &  0.000  &  0.019  \\
990517  &  22.389  &  -0.090  &  0.060  &  0.023  &  22.175  &  -0.032  &  0.000  &  0.034  &  21.495  &  -0.043  &  0.000  &  0.026  \\
990519  &  22.430  &  -0.090  &  0.060  &  0.023  &  22.173  &  -0.031  &  0.000  &  0.029  &  21.494  &  -0.083  &  0.000  &  0.036  \\
990520  &  22.425  &  -0.088  &  0.060  &  0.041  &  22.149  &  -0.007  &  0.000  &  0.023  &  21.472  &  -0.091  &  0.000  &  0.020  \\
990926  &  22.204  &  -0.047  &  0.060  &  0.037  &  22.030  &  -0.006  &  0.000  &  0.040  &  21.342  &  -0.100  &  0.000  &  0.071  \\
990927  &  22.208  &  -0.147  &  0.060  &  0.016  &  22.045  &  -0.050  &  0.000  &  0.014  &  21.363  &  -0.108  &  0.000  &  0.031  \\
990928  &  22.211  &  -0.130  &  0.060  &  0.030  &  22.054  &  -0.064  &  0.000  &  0.019  &  21.392  &  -0.105  &  0.000  &  0.064  \\
990929  &  22.240  &  -0.125  &  0.060  &  0.028  &  22.072  &  -0.073  &  0.000  &  0.021  &  21.381  &  -0.110  &  0.000  &  0.026  \\
\enddata
\tablecomments{Photometric solutions for all of the clear nights. Column~1 
lists the UT date of the observations, while columns 2 -- 4, 6 -- 8, 
and 10 -- 12 list the coefficients of the photometric solutions defined in 
\S\ref{sec:stand}. Columns 5, 9, and 13 list the 
rms scatter in the solution for each filter and each night. 
}
\end{deluxetable}
\end{center}

\begin{center}
\begin{deluxetable}{cccccc}
\tablenum{3}
\tablewidth{0pt}
\tablecaption{Differential $J$ Number Counts \label{tbl:numj} }
\tablehead {
  \colhead{Mag} &
  \colhead{$N_{raw}$} &
  \colhead{Area} &
  \colhead{$N_{corr}$} &
  \colhead{$N_{low}$} &
  \colhead{$N_{upp}$} \\
}
\startdata
15.50 . . . . . . . . . . &   2 & 0.05150 &   117 &   53 &  230 \\
16.00 . . . . . . . . . . &   2 & 0.05150 &    78 &   28 &  180 \\
16.50 . . . . . . . . . . &   4 & 0.05150 &   155 &   81 &  278 \\
17.00 . . . . . . . . . . &   3 & 0.05150 &   272 &  172 &  418 \\
17.50 . . . . . . . . . . &  28 & 0.05150 &  1049 &  848 & 1292 \\
18.00 . . . . . . . . . . &  29 & 0.05037 &  1390 & 1156 & 1666 \\
18.50 . . . . . . . . . . &  64 & 0.04453 &  3144 & 2769 & 3566 \\
19.00 . . . . . . . . . . & 114 & 0.04022 &  5868 & 5329 & 6460 \\
19.50 . . . . . . . . . . & 145 & 0.03041 & 10261 & 9440 &11150 \\
20.00 . . . . . . . . . . & 138 & 0.02356 & 12394 &11370 &13507 \\
20.50 . . . . . . . . . . &  59 & 0.00753 & 15137 &13139 &17419 \\
\enddata
\tablecomments{
Differential $J$ number counts of galaxies. Column 1 lists the center of 
each 0.5 magnitude bin, while column 2 lists the raw number of galaxies 
detected in that magnitude range. Column 3 lists the 
total effective area of that magnitude range in square degrees and column 4 
contains the number of galaxies per magnitude per square degree after 
accounting for the aperture correction and detection probability. 
Columns 5 and 6 show the $1\sigma$ lower and upper confidence limits, 
respectively, computed as in \citet{gehrels86}. 
}
\end{deluxetable}
\end{center}

\begin{center}
\begin{deluxetable}{cccccc}
\tablenum{4}
\tablewidth{0pt}
\tablecaption{Differential $H$ Number Counts \label{tbl:numh} }
\tablehead {  
  \colhead{Mag} &
  \colhead{$N_{raw}$} &  
  \colhead{Area} &
  \colhead{$N_{corr}$} &  
  \colhead{$N_{low}$} &
  \colhead{$N_{upp}$} \\
}
\startdata
15.00 . . . . . . . . . . &   2 & 0.05310 &   75 &   27 &  175 \\
15.50 . . . . . . . . . . &   4 & 0.05310 &  151 &   79 &  270 \\
16.00 . . . . . . . . . . &   3 & 0.05310 &  113 &   52 &  223 \\
16.50 . . . . . . . . . . &  12 & 0.05310 &  791 &  620 & 1004 \\
17.00 . . . . . . . . . . &  28 & 0.05226 & 1301 & 1079 & 1565 \\
17.50 . . . . . . . . . . &  59 & 0.05145 & 2060 & 1778 & 2384 \\
18.00 . . . . . . . . . . &  97 & 0.05132 & 4131 & 3730 & 4572 \\
18.50 . . . . . . . . . . & 130 & 0.04363 & 6463 & 5919 & 7054 \\
19.00 . . . . . . . . . . & 122 & 0.02187 &11524 &10499 &12644 \\
\enddata
\tablecomments{
Same as Table~\ref{tbl:numj} for the differential $H$ number counts. 
}
\end{deluxetable}
\end{center}

\begin{center}
\begin{deluxetable}{cccccc}
\tablenum{5}
\tablewidth{0pt}
\tablecaption{Differential $K$ Number Counts \label{tbl:numk} }
\tablehead {  
  \colhead{Mag} &
  \colhead{$N_{raw}$} &  
  \colhead{Area} &
  \colhead{$N_{corr}$} &  
  \colhead{$N_{low}$} &
  \colhead{$N_{upp}$} \\
}
\startdata
14.00 . . . . . . . . . . &   2 & 0.05310 &   75 &   27 &  175 \\
14.50 . . . . . . . . . . &   3 & 0.05310 &  113 &   52 &  223 \\
15.00 . . . . . . . . . . &   4 & 0.05310 &  113 &   52 &  223 \\
15.50 . . . . . . . . . . &   5 & 0.05310 &  339 &  228 &  494 \\
16.00 . . . . . . . . . . &  21 & 0.05248 & 1029 &  832 & 1268 \\
16.50 . . . . . . . . . . &  50 & 0.05163 & 2169 & 1880 & 2499 \\
17.00 . . . . . . . . . . &  59 & 0.04617 & 3379 & 2997 & 3807 \\
17.50 . . . . . . . . . . &  81 & 0.03438 & 3607 & 3150 & 4126 \\
18.00 . . . . . . . . . . &  50 & 0.01424 & 7163 & 6163 & 8312 \\
\enddata
\tablecomments{
Same as Table~\ref{tbl:numj} for the differential $K$ number counts. 
}
\end{deluxetable}
\end{center}

\end{document}